\documentclass{article}
\usepackage{amsmath,amssymb, theorem,graphicx}
\newtheorem{lemma}{Lemma}
\newtheorem{prop}{Proposition}
\newtheorem{thm}{Theorem}
\newtheorem{coro}{Corollary}

\theorembodyfont{\rm}
\newtheorem{ex}{Example}
\newtheorem{rem}{Remark}

\def\Ae {\mathcal A}
\def\Be {\mathcal B}
\def\Ce {\mathcal C}
\def\Ha {\mathcal H}
\def\Ka {\mathcal K}

\def\Me {\mathcal M}

\def\Te {\mathcal T}

\def\Se {\mathfrak S}
\def\supp{\mathrm{supp}}
\def\<{\langle}
\def\>{\rangle}
\def\Tr{{\rm Tr}\,}
\def\ptr{{\rm Tr}}

\def\Lin{\mathrm{Lin}}

\def\rank{\mathrm{rank}}

\def\qed{{\hfill $\square$}\vskip 5pt}

\def\<{\langle}
\def\>{\rangle}

\title{Extremality conditions for generalized channels}
\author{Anna Jen\v cov\'a\footnote{Supported by by the grants 
VEGA 2/0032/09 and meta-QUTE ITMS 26240120022.}\\
{\small Mathematical Institute, Slovak Academy of Sciences,}\\
{\small \v {S}tef\'{a}nikova 49, 814 73 Bratislava, Slovakia} \\
{\small  jenca@mat.savba.sk}
}
\date{}

\begin{document}

\maketitle

\textbf{Abstract.} A generalized channel is a completely positive map that preserves trace on a given subspace. We find conditions under which a generalized 
channel with  
respect to a positively generated subspace $J$ is an extreme point in the set of 
all such generalized channels. As a special case, this yields extremality 
conditions for quantum protocols. In particular, we obtain new extremality 
conditions for quantum 1-testers with 2 outcomes, which correspond to yes/no
measurements on the set of quantum channels.

\section{Introduction}

Generalized channels with respect to a positively generated
subspace $J$ of a finite-dimensional $C^*$-algebra $\Ae$ were introduced in
\cite{ja}. These are  completely positive maps  $\Ae\to \Be$ that
preserve trace on $J$. If restricted to $J$, a generalized channel defines a
channel on $J$, that is a completely positive trace preserving map $J\to \Be$.
Conversely,
 any channel on $J$ can be extended to a generalized channel.
Since, in general, there are many generalized channels restricting to the same
map on $J$, a channel on $J$ can be viewed as an equivalence class of generalized channels.

The motivation for the study of generalized channels comes from the description
of quantum protocols by so-called generalized quantum instruments, \cite{daria,
 daria_extr}
(or quantum strategies \cite{guwat}). These include quantum combs \cite{daria_circ} describing
quantum networks, and quantum testers \cite{daria_testers}
that describe measurements on combs.
 It can be seen \cite{ja} that all generalized quantum instruments are (multiples of)  generalized channels with respect to certain subspaces that are given by the  Choi isomorphism between completely positive  maps and positive
elements in tensor products.

A particular case of a generalized channel is a generalized POVM, which describes
a measurement on the
intersection of $J$ and the state space, in the same way that  a positive
operator valued measure (POVM) describes a usual measurement.
Here, a measurement on a convex subset $K$ of the state space is naturally defined as
 an affine map from $K$ to the set of probability measures on the set of
outcomes.
But unlike the POVM,
 the generalized POVM describing a given measurement is in general not unique,
so that again, measurements are equivalence classes of generalized POVMs.

If the subspace $J$ contains a positive invertible element, the set of generalized channels with respect to $J$ is a compact convex set, see \cite[Proposition 6]{ja}.
 Hence any element of this set is a convex
combination of its extreme points. Moreover, many optimalization
problems consist in maximizing a convex function over the set of generalized
channels, and this maximum is attained at an extreme point. Therefore, it is
important to characterize  extremal generalized channels.
For generalized quantum instruments, this was done recently in \cite{daria_extr}.

In the  present paper, we obtain a set of conditions characterizing extremal
generalized channels with respect to $J$. These conditions are given in terms of
 the Choi representation, the Kraus representation and the conjugate map of 
$\Phi$.  In the case of generalized quantum instruments, we get extremality conditions different from the ones
obtained in \cite{daria_extr}. Moreover, we obtain conditions for extremality in the class of generalized POVMs for projection valued measures. In particular, 
this leads to a new
characterization of 
extremal 1-testers with two outcomes.

\section{Preliminaries}

Let $\Ae$ be a  finite dimensional $C^*$-algebra. Then $\Ae$ is
isomorphic to a direct sum of matrix algebras, that is, there are finite
dimensional Hilbert spaces $\Ha_1,\dots\Ha_n$, such that
\[
\Ae\equiv \bigoplus_j B(\Ha_j)
\]
Below we always assume that $\Ae$ is equal to this direct sum. Let $s_1,\dots s_l$ be the minimal central projections in $\Ae$ and let us fix an orthonormal basis  $|i\>$, $i=1,\dots, N$  of $\Ha=\oplus_j\Ha_j$, such that each $|i\>\<i|$ commutes with all $s_j$. Then $\Ae$ is identified with the  subalgebra of block-diagonal matrices in  the matrix 
algebra $M_N(\mathbb C)\equiv B(\Ha)$.  The identity in $\Ae$ will be denoted by 
$I_\Ae$.

We fix a trace $\ptr_\Ae$ on $\Ae$ to be the restriction of the trace $\ptr_\Ha$
in $B(\Ha)$, we omit the subscript
$\Ae$ if no confusion is possible. The trace defines the Hilbert-Schmidt inner product in $\Ae$ by $\<a,b\>=\Tr a^*b$. If $A\subset \Ae$ we denote by $A^\perp$ the orthogonal complement of $A$ with respect to $\<\cdot,\cdot\>$.
 Let $E_\Ae: B(\Ha)\to \Ae$ be defined as 
\[
E_\Ae(a)=\sum_i s_ias_i,\qquad a\in B(\Ha).
\]
 Then $\Tr ab=\Tr E_\Ae(a)b$ for all $a\in B(\Ha)$, $b\in \Ae$ and  $E_\Ae$ is the trace-preserving conditional expectation onto $\Ae$.

If $\Be$ is another $C^*$ algebra, then $\ptr^{\Ae\otimes\Be}_\Ae$ will
denote the partial trace on the tensor product $\Ae\otimes\Be$, $\ptr^{\Ae\otimes \Be}_\Ae (a\otimes b)=\Tr( a)b$. If the input space is clear, we will
denote the partial trace just by $\ptr_\Ae$.

For $a\in \Ae$, we denote by $a^T$ the transpose of $a$. Note that $\ptr^{\Ae\otimes \Be}_\Ae (x^T)=(\ptr^{\Ae\otimes\Be}_\Ae x)^T$ for $x\in \Ae\otimes \Be$.
If $A\subset \Ae$, then $A^T=\{a^T,\ a\in A\}$.

We denote by $\Ae^+$ the
convex cone  of positive elements in $\Ae$ and $\Se(\Ae)$ the set of states on
$\Ae$, which will be identified with the set of density operators in $\Ae$,
that is, elements  $\rho\in \Ae^+$ with $\Tr\rho=1$.
For $a\in \Ae^+$, the projection onto the support
of $a$ will be denoted by $\supp(a)$. If $p\in \Ae$ is a projection, we denote $\Ae_p=p\Ae p$.

Let  $L\subseteq \Ae$ be a linear subspace, then $L$ is self-adjoint if
$a^*\in L$ whenever $a\in L$. 
If $L$ is generated by positive elements, then we say that $L$ is positively generated. It is clear that $L$ is self-adjoint in this case.

\subsection{Completely positive maps, channels and generalized channels}

Let $\Ae\subseteq B(\Ha)$, $\Be\subseteq B(\Ka)$  be   finite dimensional
$C^*$ algebras and let
$J\subseteq \Ae$   be a positively generated subspace of $\Ae$.
A linear map $\Xi:J\to \Be$ is positive if $\Xi$ maps $J\cap \Ae^+$ into $\Be^+$
 and it is completely positive  if the map
\[
\Xi\otimes id_{\Ha_0} : J\otimes B(\Ha_0)\to \Be\otimes B(\Ha_0)
\]
is positive for any finite dimensional Hilbert space $\Ha_0$. In this case,
$\Xi$ extends to a completely positive  map $\Phi:\Ae\to \Be$ \cite{ja}.

Let $\Phi:\Ae\to \Be$ be a linear map.  Suppose first that $\Ae=B(\Ha)$, then 
the Choi representation of $\Phi$ is defined as
\[
X_\Phi=(\Phi\otimes id_\Ha)(|\psi_\Ha\>\<\psi_\Ha|),
\]
where $|\psi_\Ha\>=\sum_i|i\>\otimes |i\>$, note that  
\begin{equation}\label{eq:psi}
(a\otimes I_\Ha)|\psi_\Ha\>=(I_\Ha\otimes a^T)|\psi_\Ha\>, \quad a\in B(\Ha).
\end{equation}
We have 
\begin{equation}\label{eq:choi_map}
\Phi(a)= \ptr_\Ae[(I_\Be\otimes a^T)X_\Phi],\qquad a\in B(\Ha)
\end{equation}
and $\Phi$ is completely positive if and only if $X_\Phi\ge 0$, \cite{choi}. 
Let now $\Ae\subset B(\Ha)$ and let  $\Phi'=\Phi\circ E_\Ae$. Then $\Phi'$ is an  extension of $\Phi$ and it is completely positive  if and only if $\Phi$ is.  
In this case, we define the Choi representation of $\Phi$ as
\[
X_\Phi:=X_{\Phi'}.
\]
It is easy to see that $X_\Phi\in \Ae\otimes \Be$ and 
that $\Phi$ is completely positive if and only if $X_\Phi\ge 0$. Moreover, 
 (\ref{eq:choi_map}) holds 
if $a\in \Ae$. If $\Ae=\bigoplus_jB(\Ha_j)$, $\Be=\bigoplus_k B(\Ka_k)$, then 
there 
are  maps $\Phi_{jk}:B(\Ha_j)\to B(\Ka_k)$ such that
\begin{equation}\label{eq:mapsdec}
\Phi(a)=\oplus_k \Phi_{jk}(a),\qquad a\in B(\Ha_j)
\end{equation}
It is clear that $\Phi$ is completely positive if and only if $\Phi_{jk}$ are 
completely positive for all $j,k$.

A channel $\Xi:J\to \Be$ is a trace preserving completely positive  map. Clearly, a completely positive  map
 $\Phi:\Ae \to \Be$ is an extension of some channel $J\to \Be$ if and only if $\Phi$ preserves trace on $J$, such a map is called a
generalized channel. If $X_\Phi\in (\Be\otimes \Ae)^+$ is the Choi representation 
of $\Phi:\Ae\to \Be $, then
$\Phi$ is a generalized channel if and only if
\[
\ptr_\Be X\in  (I_\Ae+(J^T)^\perp)\cap \Ae^+
\]
where $J^\perp$ denotes the orthogonal complement with respect to the Hilbert-Schmidt inner product.
Let us denote the set of all generalized channels, resp. their Choi matrices, by $\Ce_J(\Ae,\Be)$. Clearly, if $J=\Ae$, we obtain the set of usual channels $\Ae\to \Be$, this set will be denoted by $\Ce(\Ae,\Be)$.

Let $c\in\Ae$ and let us denote $\chi_c(a)=cac^*$. Then $\chi_c:\Ae\to \Ae$ is  
clearly completely positive and it is a generalized channel if and only if
\[
c^*c\in (I_\Ae+ J^\perp)\cap \Ae^+,
\]
 such generalized channels are called simple. The following is a slight modification of \cite[Proposition 5]{ja}.

 \begin{prop}
 Let $\Phi:\Ae\to \Be$ be a linear map. Then $\Phi$ is a generalized channel with respect to $J$ if and only if there is a simple generalized channel $\chi_c:\Ae\to \Ae$ and a channel
 $\Lambda:\Ae\to \Be$, such that
 \begin{equation}\label{eq:decomp}
 \Phi=\Lambda\circ \chi_c.
 \end{equation}
 Moreover, let $\Lambda_q$ be the restriction of $\Lambda$ to $\Ae_q$,  $q=\supp(cc^*)$, then the  pairs $(\Lambda,c)$ and $(\Lambda',c')$ define the same generalized channel if and only if there is a partial isometry $V$ with $V^*V=q$, $VV^*=q'=\supp (c'c^{'*})$ such that $c'=Vc$ and $\Lambda'_{q'}=\Lambda_q\circ \mathrm{Ad}_{V^*}$.

  \end{prop}

The decomposition $\Phi=\Lambda_q\circ \chi_c$ in the above proposition will be
 called minimal.
Note that in terms of the Choi matrices, (\ref{eq:decomp})
has the form
\[
X_\Phi=(I_\Be\otimes c^T)X_\Lambda(I_\Be\otimes (c^T)^*).
  \]
Note also that $c^*c=\Phi^*(I_\Be)=(\Tr_\Be X_\Phi)^T$ whenever (\ref{eq:decomp}) holds.

Important examples of generalized channels will be treated in the sections 
below.  In all examples, the subspace $J$ is of the form
 $J=S^{-1}(J_0)$, where $S:\Ae\to \Ae_0$ is a channel and $J_0\subseteq \Ae_0$ a 
subspace. Then \cite{ja}
\begin{equation} \label{eq:perp}
J^\perp=S^*(J_0^\perp),
\end{equation}
where $S^*:\Ae_0\to\Ae$ is the adjoint of $S$.
Let $J_0$ be the one dimensional subspace  generated by some $\rho_0=S(\rho)$ with
invertible $\rho\in \Se(\Ae)$, then 
\begin{equation}\label{eq:hom}
J\cap \Se(\Ae)=\{\sigma\in \Se(\Ae),\ S(\sigma)=\rho_0\}=: K_{S,\rho_0}
\end{equation}
is the equivalence class containing $\rho$ of the obvious equivalence relation on
$\Ae$ defined by $S$. 

Suppose  moreover that the adjoint $S^*$ is an injective homomorphism, that is, 
$S^*(a_0b_0)=S^*(a_0)S^*(b_0)$ for all $a_0,b_0\in \Ae_0$ and $S^*(a_0)=0$ implies $a_0=0$ for $a_0\in \Ae_0$, note that this implies that  $\rho_0$ is invertible.
 We will describe the simple generalized channels in this case.

\begin{lemma}\label{lemma:hom} Let $J$ be as above, then  
$d\in (I_\Ae+J^\perp)\cap \Ae^+$ if and only if $d=S^*(b^2_0)$, where 
$b_0=\sigma_0^{1/2}(\sigma_0^{1/2}\rho_0\sigma_0^{1/2})^{-1/2}\sigma_0^{1/2}$ 
for some $\sigma_0\in \Se(\Ae_0)$. Moreover, in this case,
\[
\chi_{d^{1/2}}(K_{S,\rho_0})\subseteq K_{S,\sigma_0}
\]

\end{lemma}

{\it Proof.}  By (\ref{eq:perp}), we obtain 
\[
(I_\Ae+J^\perp)\cap \Ae^+=(I_{\Ae}+S^*(\{\rho_0\}^\perp))\cap \Ae^+=
S^*((I_{\Ae_0}+\{\rho_0\}^\perp)\cap \Ae_0^+)
\]
Moreover, let $b_0\in \Ae_0^+$ and put $\sigma_0:=b_0\rho_0b_0$, then 
$b_0=\sigma_0^{1/2}(\sigma_0^{1/2}\rho_0\sigma_0^{1/2})^{-1/2}\sigma_0^{1/2}$ is 
the unique positive solution of this equation. Clearly,
$b_0^2\in I_{\Ae_0}+\{\rho_0\}^\perp$ if and only if $\sigma_0 \in \Se(\Ae_0)$.

Let now $\chi_{d^{1/2}}$ be a simple generalized channel, so that $d^{1/2}=
S^*(b_0)$. For $a\in \Ae$, $a_0\in \Ae_0$, we have
\[
\Tr S(\chi_{d^{1/2}}(a))a_0=\Tr S^*(b_0)aS^*(b_0)S^*(a_0)=\Tr b_0S(a)b_0a_0
\]
so that $S(\chi_{d^{1/2}}(a))=\chi_{b_0}(S(a))=\sigma_0$ if $S(a)=\rho_0$.

\qed

An example of this situation is contained in Section \ref{sec:chanonchan}. 
We  will show one more simple example.

\begin{ex}\label{ex:diag} 
Let $\mathcal D_n$ denote the set of density matrices in the matrix algebra 
$M_n(\mathbb C)$.
Let $\lambda=(\lambda_1,\dots,\lambda_n)$, $\lambda_i>0$, $\sum_i\lambda_i=1$ and let $Diag_\lambda$ be the set of density matrices $\rho=\{\rho_{ij}\}_{i,j}\in \mathcal D_n$ such that $\rho_{ii}=\lambda_i$ for all 
$i$.  Let $S:M_n(\mathbb C)\to \mathbb C^n$ be the map 
that maps every matrix onto the vector of its diagonal elements, that is,
\[
S(a)=(\<1,a1\>,\dots,\<n, an\>),\qquad a\in M_n(\mathbb C)
\]
where $|i\>$ denotes the standard  basis of $\mathbb C^n$.
Then $S$ is a channel and $Diag_\lambda= J_\lambda\cap \mathcal D_n=K_{S,\lambda}$
in the notation of (\ref{eq:hom}), here
$J_\lambda:=S^{-1}(\mathbb C \lambda)$. Moreover, it is easy to see that 
$S^*(x)=\sum_ix_i|i\>\<i|$ for any $x\in \mathbb C^n$ is an isomorphism onto the 
commutative subalgebra generated by $|i\>\<i|$. It follows that all simple 
generalized channels are obtained from elements of the form $c=Ud^{1/2}$, where 
$U$ is a partial isometry and 
$d=\sum_i d_i|i\>\<i|$, $d_i=\mu_i/\lambda_i$ for some probability vector 
$\mu=(\mu_1,\dots,\mu_n)$.
In particular, 
 $\chi_{d^{1/2}}$ maps $Diag_\lambda$ onto $Diag_\mu$.

\end{ex}

\subsubsection{Channels on channels} \label{sec:chanonchan}
 Let $\Ha_0,\Ha_1$ be finite dimensional
Hilbert spaces and let $\Ce(\Ha_0,\Ha_1)$ denote the set of Choi matrices of
channels $B(\Ha_0)\to B(\Ha_1)$. Then
\[
\Ce(\Ha_0,\Ha_1)=\{ X\in B(\Ha_1\otimes\Ha_0)^+, \ptr_{\Ha_1} X=I_{\Ha_0}\}
\]
Let $\Ae=B(\Ha_1\otimes \Ha_0)$ and let
$J\subset\Ae$ be the subspace generated by $\Ce(\Ha_0,\Ha_1)$. Then
$J= \ptr_{\Ha_1}^{-1}(\mathbb C I_{\Ha_0})$ and
$\Ce(\Ha_0,\Ha_1)= J\cap \dim(\Ha_0)\Se(\Ae)$, hence $\Ce(\Ha_0,\Ha_1)$ is a 
constant multiple of a set of the form (\ref{eq:hom}). Moreover, 
\[
S^*(a_0)=\ptr_{\Ha_1}^*(a_0)=I_{\Ha_1}\otimes a_0, \qquad a_0\in B(\Ha_0),
\]
 hence
 we may apply Lemma \ref{lemma:hom}. Note also that  $J^T=J$ and
\[
J^\perp=I_{\Ha_1}\otimes \Te(\Ha_0),
\]
where $\Te(\Ha_0)$ is the subspace of traceless elements in $B(\Ha_0)$.

Let $\Be$ be a finite dimensional $C^*$ algebra.
Let us denote by $\Ce(\Ha_0,\Ha_1,\Be)$ the set of Choi matrices of completely 
positive maps
$\Ae\to \Be$ that map $\Ce(\Ha_0,\Ha_1)$ into the state space $\Se(\Be)$.
Then $\Ce(\Ha_0,\Ha_1,\Be)=\dim(\Ha_0)^{-1} \Ce_J(\Ae,\Be)$.
It follows that $X\in (\Be\otimes \Ae)^+$ is in $\Ce(\Ha_0,\Ha_1,\Be)$ if and only if
\[
 \ptr_\Be X=I_{\Ha_1}\otimes \omega,\qquad  \omega\in \Se(\Ha_0)
\]

Let $X=X_\Phi\in  \Ce(\Ha_0,\Ha_1,\Be)$ and let 
$\ptr_\Be X=I_{\Ha_1}\otimes \omega$, $\omega\in B(\Ha_0)$. Let
\[
\Phi=\Lambda_q\circ\chi_{I\otimes c}, \qquad c^*c=\omega, \quad q=\supp(cc^*)
\]
  be a minimal decomposition.  Then for any $X_{\mathcal E}\in \Ce(\Ha_0,\Ha_1)$, we have
\[
\chi_{I\otimes c}(X_{\mathcal E})=\chi_{I\otimes c}\circ(\mathcal E\otimes id_{\Ha_0})(|\psi_{\Ha_0}\>\<\psi_{\Ha_0}|)=
(\mathcal E\otimes id_{q\Ha_0})(\rho)
\]
where $\rho=\chi_{I\otimes c}(|\psi_{\Ha_0}\>\<\psi_{\Ha_0}|)\in B(\Ha_0\otimes q\Ha_0)$ is a pure state, such that
$\ptr_2\rho=\omega^T$. This leads to the following implementation for $\Phi$ 
(\cite[Theorem 4]{ja}):

\begin{prop}
 There is a Hilbert space $\Ha_A$, a
 pure state $\rho\in \Ha_0\otimes\Ha_A$ and a channel $\Lambda: B(\Ha_1\otimes \Ha_A)\to \Be$ such that
\begin{equation}\label{eq:imp}
\Phi(X_{\mathcal E})=\Lambda\circ(\mathcal E\otimes id_{\Ha_A})(\rho),
\end{equation}
for all channels $\mathcal E:B(\Ha_0)\to B(\Ha_1)$. Conversely, any map of the form (\ref{eq:imp}) defines an element in $\Ce(\Ha_0,\Ha_1,\Be)$.
\end{prop}

The triple $(\Ha_A,\rho,\Lambda)$ as above will be  called a representation of $X$.
 Moreover, if  $\dim(\Ha_A)=\mathrm{rank}(\ptr_{\Ha_A} \rho)$, then the triple  $(\Ha_A,\rho,\Lambda)$ will be called a minimal representation of $X$. If $(\Ha_A,\rho,\Lambda)$ and $(\Ha_B,\rho',\Lambda')$ are two minimal representations of $X$, then there is a unitary operator $U:\Ha_A\to \Ha_B$ such that 
$\rho'=(I\otimes U)\rho (I\otimes U^*)$ and $\Lambda'=\Lambda\circ Ad_{I\otimes U^*}$.

\subsubsection{Quantum supermaps and generalized quantum instruments}
\label{sec:supermaps}
Let $\Be_0,\Be_1,\dots,$ be a sequence of finite dimensional $C^*$-algebras
 and let $\Ae_n:=\Be_n\otimes\dots\otimes\Be_0$.
For  $n\in \mathbb N$, the sets $\Ce(\Be_0,\dots,\Be_n)$ are defined as
follows:

Let $\Ce(\Be_0,\Be_1)$ be the set of the Choi matrices of channels
$\Be_0\to \Be_1$. For $n>1$, $\Ce(\Be_0,\dots,\Be_n)$ is defined as the set of
Choi matrices of
completely positive maps $\Ae_{n-1}\to \Be_n$, that map $\Ce(\Be_0,\dots,\Be_{n-1})$ into
$\Se(\Be_n)$.  Such maps are called quantum supermaps, note that this definition is different from the
definition of a quantum supermap given in \cite{daria_super}. It is clear that $\Ce(\Ha_0,
\Ha_1,\Be)$ from Section \ref{sec:chanonchan} is a set of quantum supermaps for $n=2$.

The set  $\Ce(\Be_0,\dots,\Be_n)$ is a constant multiple of a set of generalized channels, more precisely,
\[
 \Ce(\Be_0,\dots,\Be_n)= J_n\cap c_n \Se(\Ae_n) =
\frac1{c_{n-1}}\Ce_{J_{n-1}}(\Ae_{n-1},\Be_n)
\]
where $c_n:=\Pi_{l=0}^{\lfloor \frac{n-1}2\rfloor}
\Tr(I_{\Be_{n-1-2l}})$ and $J_n:=J_n(\Be_0,\dots,\Be_n)$ denotes the subspace in
$\Ae_n$ generated by
$\Ce(\Be_0,\dots,\Be_n)$. The subspaces $J_n$ are clearly positively generated and
obtained as follows:
Let $S_n :\Ae_n\to \Ae_{n-1}$ denote the partial trace
$\ptr^{\Ae_n}_{\Be_n}$, $n=1,2,\dots$, then we have
\begin{eqnarray}
J_{2k-1} &=& S_{2k-1}^{-1}(S_{2k-2}^*(S_{2k-3}^{-1}(\dots S_1^{-1}(\mathbb C
I_{\Be_0})\dots)))\label{eq:jodd}\\
J_{2k} &=& S_{2k}^{-1}(S_{2k-1}^*(S_{2k-2}^{-1}(\dots S_1^*(\Be_0)\dots)))\label{eq:jeven}
\end{eqnarray}
There is another way to characterize the elements in $\Ce(\Be_0,\dots,\Be_n)$:
Let  $k:=\lfloor \frac n2\rfloor$.
Then $X\in \Ce(\Be_0,\dots,\Be_n)$ if and only if there are positive
elements $Y^{(m)}\in \Ae_{n-2m}$ for
$m=0,\dots,k$, such that
\begin{equation}\label{eq:daria}
\ptr_{\Be_{n-2m}} Y^{(m)}=I_{\Be_{n-2m-1}}\otimes Y^{(m+1)}, m=0,\dots,k-1
\end{equation}
 $Y^{(0)}:=X$,  $Y^{(k)} \in \Ce(\Be_0,\Be_1)$ if $n=2k+1$
and $Y^{(k)}\in \Se(\Be_0)$ if $n=2k$.

\begin{figure}
\begin{center}
\includegraphics{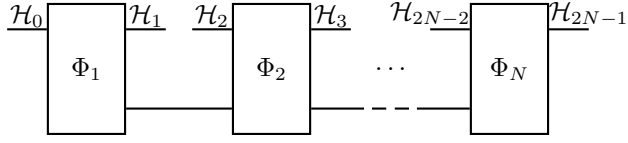}
\caption{A deterministic quantum $N$-comb}\label{fig:comb}

\end{center}
\end{figure}

Let $\Ha_0,\Ha_1,\dots$ be finite dimensional Hilbert spaces.
As it was pointed out in \cite{ja}, the elements of
$\Ce(\Ha_0,\dots,\Ha_{2N-1})$ are precisely the (deterministic) quantum $N$-combs 
on $(\Ha_0,\dots,\Ha_{2N-1})$
 \cite{daria}, which are the Choi matrices of completely positive maps used in description of quantum networks. More precisely, quantum 1-combs are defined as the elements of
$\Ce(\Ha_0,\Ha_1)$ and for $N>1$, quantum $N$-combs are completely positive maps 
that map the $N-1$ combs on $(\Ha_1,\dots,\Ha_{2N-1})$ to $\Ce(\Ha_0,\Ha_{2N-1})$.
Quantum $N$-combs can be represented by memory channels, given  by a sequence of 
$N$ channels connected by an ancilla, see Fig. \ref{fig:comb}

Another special case is the set of quantum  $N$-testers (with $m$ outcomes), which describe measurements on quantum $N$-combs. These  are 
precisely the elements of the set
$\Ce(\Ha_0,\dots,\Ha_{2N-1},\mathbb C^m)$. More generally, if some of the channels in the sequence representing the  $N$-comb are replaced by instruments or POVMs, we obtain the set of generalized quantum instruments, \cite{daria, daria_extr}. These are described by probabilistic $N$-combs, which are positive operators $X_i$, such that $\sum_iX_i$ is a (deterministic) quantum $N$-comb.

\begin{figure}
\begin{center}
\includegraphics{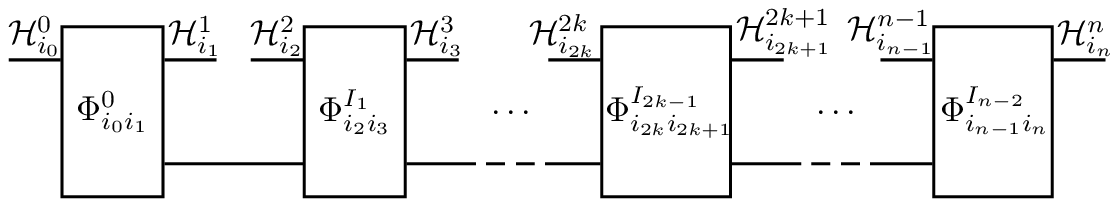}
\caption{Representation of a  quantum supermap, $n=2N-1$, $I=(i_0,\dots,i_n)$}\label{fig:condc}

\end{center}
\end{figure}

\begin{figure}
\begin{center}
\includegraphics{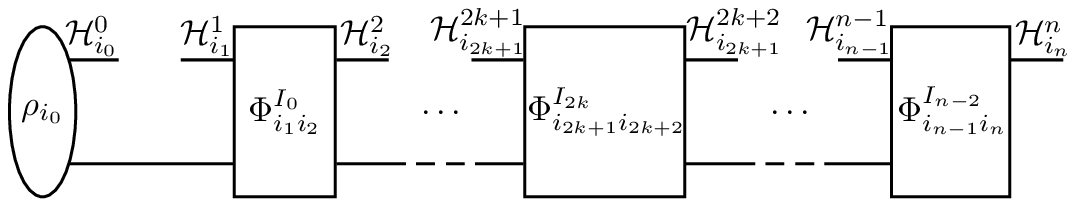}
\caption{Representation of a quantum supermap, $n=2N$, $I=(i_0,\dots,i_n)$}\label{fig:supm}

\end{center}
\end{figure}

If the matrix algebras $B(\Ha_n)$ are replaced by finite dimensional $C^*$-algebras, it can be seen that the set of quantum supermaps with $n=2N-1$
correspond to conditional quantum combs \cite{daria_bit}, which describe 
 quantum protocols where classical inputs and outputs are
 allowed.  Let $\Be_j=\bigoplus_{k=1}^{n_j} B(\Ha^j_k)$, $j=0,\dots,n$, then 
any element in $\Ce(\Be_0,\dots,\Be_{2N-1})$ is represented by a sequence of 
networks as in Fig. \ref{fig:condc}. The elements of the sequence are labelled by multiindices $I=(i_0,\dots,i_n)$, $i_j\in\{1,\dots,n_j\}$, representing the 
classical inputs and outputs, $I_j=(i_0,\dots,i_j)$ and $\Phi^0: \Be_0\to \Be_1$, $\Phi^{I_{2k-1}}: 
\Be_{2k}\to\Be_{2k+1}$, $k=1,\dots, N-1$ are channels.
For $n=2N$, the elements in $\Ce(\Be_0,\dots,\Be_n)$ are represented similarly, 
see Fig. \ref{fig:supm}, here $\rho_0=\oplus_{i_0}\rho_{i_0}\in \Se(\Be_0)$ and $\Phi^{I_{2k}}:\Be_{2k+1}\to\Be_{2k+2}$ are channels, $k=0,\dots,N-1$, see 
\cite[Theorem 8]{ja}.

\section{Extremal generalized channels}\label{sec:extr_gchannels}

Let $J\subseteq \Ae$ be a positively generated subspace.
Let us denote
\[
L:=\{ X\in \Be\otimes\Ae, \ptr_\Be(X)\in I_\Ae+(J^T)^\perp\},
\]
then $L$ is an affine subspace in  $\Be\otimes\Ae$ and we have
\begin{equation}\label{eq:Lin}
\Lin(L):= \{ X-Y: X,Y\in L\}=\{X\in \Be\otimes\Ae, \ptr_\Be(X)\in (J^T)^\perp\}
\end{equation}
 It is easy to see that
\begin{equation}\label{eq:gchanJ}
\Ce_J(\Ae,\Be)=(\Be\otimes\Ae)^+\cap L,
\end{equation}
It is also clear that the faces in $\Ce_J(\Ae,\Be)$ are precisely the sets of the form
\[
F=(\Be\otimes\Ae)^+_P\cap L
 \]
 for a projection $P\in \Be\otimes\Ae$. Indeed, it is quite clear that any such subset is a face. Conversely,
let $F\subset \Ce_J(\Ae,\Be)$ be a face and let $X\in F$ be an element with maximal support, $P=\supp(X)$.
Then $F\subseteq G:=(\Be\otimes\Ae)^+_P\cap L$. Since both $F$ and $G$ are faces in $\Ce_J(\Ae,\Be)$ and
$X\in F\cap ri(G)$, where $ri(G)$ denotes the relative interior of $G$, it follows by \cite[Theorem, 18.1]{rockafellar} that $F=G$.

\begin{thm}\label{thm:extreme_gchan} Let $X\in \Ce_J(\Ae,\Be)$ and let $P=\supp(X)$.
Then $X$ is extremal  if and only if
\[
(\Be\otimes \Ae)_P\cap \Lin(L)=\{0\}
\]
\end{thm}

{\it Proof.} Suppose that $X=\frac12 (X_1+X_2)$ for some $X_1,X_2\in \Ce_J(\Ae,\Be)$. Then
$X_1,X_2\in (\Be\otimes \Ae)^+_P\cap L$ so that $X_1-X_2\in (\Ae\otimes \Be)_P\cap \Lin(L)$.
Conversely, let $Y\in (\Be\otimes \Ae)_P\cap \Lin(L)$, then also $Y+Y^*\in (\Be\otimes \Ae)_P\cap \Lin(L)$, hence we
 may suppose that $Y$ is self adjoint. Then there exists some $t>0$ such that
$Z_\pm:=X\pm tY\in \Ce_J(\Ae,\Be)$, moreover, $X=\frac12(Z_++Z_-)$.

\qed

\begin{coro}\label{coro:gchan_dim} Let $X$ be an extreme point in $\Ce_J(\Ae,\Be)$. Let
 $P=\supp(X)$ and let  $p=\supp(\ptr_\Be X)^T$. Then
\[
\dim (\Be\otimes \Ae)_P\le \dim(pJp)
\]

\end{coro}

{\it Proof.} Note that we have
\[
\Tr X(I_{\Be\otimes \Ae}-I_\Be\otimes p^T)=\Tr X(I_\Be\otimes (I_\Ae-p^T))=
\Tr (I_\Ae-p^T)\ptr_\Be X =0,
\]
 so that $P\le I_\Be\otimes p^T$.
Let
$L_{p^T}:=L\cap (\Be\otimes \Ae_{p^T})$,  then
\begin{equation}\label{eq:Lp}
(\Be\otimes \Ae)_P\cap \Lin(L_{p^T })=(\Be\otimes \Ae)_P\cap \Lin(L)
\end{equation}
Moreover,  since $(J^T)^\perp\cap\Ae_{p^T}=((pJp)^T)^\perp\cap \Ae_{p^T}$,
\[
\Lin(L_{p^T})=\{Y\in\Be\otimes\Ae_{p^T},\ \ptr_\Be Y\in ((pJp)^T)^\perp\cap \Ae_{p^T}\}.
\]
 Suppose that
 $X$ is an extreme point, then $(\Be\otimes \Ae)_P\cap \Lin(L_{p^T})=\{0\}$, so that,
 by applying the orthogonal complements in $\Be\otimes\Ae_{p^T}$,
\[(\Be\otimes \Ae)_P^\perp\vee \Lin(L_{p^T})^\perp=\Be\otimes \Ae_{p^T}.\]
It follows that we must have  $\dim((\Be\otimes \Ae)_P)\le \dim(\Lin(L_{p^T})^\perp)$.
The statement now follows from $\Lin(L_{p^T})^\perp=I_\Be\otimes (pJp)^T$, by 
(\ref{eq:perp}).

\qed

Let $\Phi=\Lambda_q\circ\chi_c$ be a minimal decomposition of $\Phi$, $p=\supp(c^*c)$, $q=\supp(cc^*)$.
 The next Proposition gives the extremality condition in terms of the channel $\Lambda_q$. Note that as a channel, 
 $\Lambda_q$ is also a generalized channel with
 respect to any subspace in $\Ae_q$.

\begin{thm}\label{thm:extremal_decomp}  $\Phi$ is  extremal in $\Ce_J(\Ae,\Be)$ if and only if
$\Lambda_q$ is extremal in $\Ce_{cJc^*}(\Ae_q,\Be)$.

\end{thm}

{\it Proof.} Suppose $\Lambda_q=\frac 12(\Psi_1+\Psi_2)$ for some $\Psi_1\neq \Psi_2\in \Ce_{cJc^*}(\Ae_q,\Be)$. Put
$\Phi_i=\Psi_i\circ\chi_c$, then $\Phi_i$ is completely positive and for any $a\in J$
\[
\Tr \Phi_i(a)=\Tr \Psi_i(cac^*)=\Tr cac^*=\Tr a
\]
so that $\Phi_i\in \Ce_J(\Ae,\Be)$. Clearly, $\Phi=\frac 12(\Phi_1+\Phi_2)$ and $\Phi_1\neq \Phi_2$.

Conversely, suppose that $\Phi=\frac 12(\Phi_1+\Phi_2)$, $\Phi_1\neq \Phi_2\in \Ce_J(\Ae,\Be)$. Then $\Phi_i$ are decomposed as
$\Phi_i=\Lambda_i\circ\chi_{c_i}$ for some channels $\Lambda_i:\Ae\to \Be$ and $\Phi_i^*(I_\Be)=c_i^*c_i$. Clearly,  
$c^*c=\frac 12(c_1^*c_1+c_2^*c_2)$, so that $\supp (c_i^*c_i)\le \supp (c^*c)=p$. Put 
$\Psi_i=\Phi_i\circ \chi_{c^{-1}}$, where $c^{-1}=(c^*c)^{-1/2}U^*$ for the polar decomposition $c=U(c^*c)^{1/2}$, the inverse 
 being taken on the support of $c^*c$. Then $\Phi_i$ is a well defined completely positive map $\Ae_q\to \Be$. Moreover,
 \[
\Tr \Psi_i(cyc^*)=\Tr \Phi_i(y)=\Tr y,\qquad y\in J
 \]
so that $\Psi_i\in \Ce_{cJc^*}(\Ae_q,\Be)$. We also have
\[
\frac12(\Psi_1+\Psi_2)\circ\chi_c=\frac 12 (\Phi_1+\Phi_2)\circ\chi_{c^{-1}}\circ\chi_c=\Phi\circ \chi_p=\Phi=\Lambda_q\circ\chi_c
\]
By uniqueness, $\Lambda_q=\frac12(\Psi_1+\Psi_2)$.

\qed

\subsection{Extremality conditions for Kraus operators}

Let $\Ae=\oplus_i B(\Ha_i)$, $\Be=\oplus_j B(\Ka_j)$, $\Ha=\oplus_i \Ha_i$,
 $\Ka=\oplus_j \Ka_j$. Let $\{|v_k\>\}_k$ be a set of vectors in $\Ka\otimes \Ha$, such that
 \[ 
 X_\Phi=\sum_k |v_k\>\<v_k|
 \]
 and for each $i,j$,  there is a set $I(i,j)$ of indices such that
$|v^k\>\in \Ka_i\otimes\Ha_j$ for $ k\in I(i,j)$. For example, we can choose $|v_k\>$ to be  eigenvectors
 of $X_\Phi$. It is quite clear that the vectors $|v_k\>$ span $P(\Ka\otimes \Ha)$, where $P=\supp(X_\Phi)$.

 For each $j$, let  $\{|\alpha_m^j\>\}_m$ be an ONB in $\Ha_j$. Then
for each $k\in I(i,j)$ there are some vectors $|v^k_m\>\in \Ka_i$, such that
$|v^k\>=\sum_m  |v^k_m\>\otimes|\alpha^j_m\>$. Moreover, there is a one-to-one
correspondence between vectors in $\Ka_i\otimes \Ha_j$ and linear maps
$\Ha_j\to\Ka_i$, given by $|x\>\otimes|y\>\mapsto |x\>\<y|$. In this correspondence,
\[
|v^k\>\mapsto V_k:=\sum_m |v^k_m\>\<\alpha^j_m|.
\]
 The  map
$\Phi$ has  the form
\begin{equation}\label{eq:minimal_kraus}
\Phi(a)=\sum_k V_kaV_k^*=\sum_{i,j}\sum_{k\in I(i,j)}V_kaV_k^*
\end{equation}
so that $V_k$ are Kraus operators of $\Phi$.

Suppose  that the vectors $\{|v_k\>\}_k$ are linearly independent, then the operators
$V_k$ are
linearly independent as well and (\ref{eq:minimal_kraus}) is a minimal Kraus representation
of $\Phi$, \cite{choi}. Conversely, any minimal Kraus representation is obtained in this way.
It follows that $\{|v^k\>\<v^l|,\ k,l\in I(i,j)\}_{i,j}$
is a basis of $(\Be\otimes \Ae)_P$, so that any element $D\in(\Be\otimes \Ae)_P$
can be uniquely expressed as
\begin{equation}\label{eq:D}
D=\sum_{i,j}\sum_{k,l\in I(i,j)} d_{k,l} |v^k\>\<v^l|
\end{equation}

Let us fix a basis $\{x_1,\dots,x_M\}$  of $J^\perp$.

\begin{coro}\label{coro:extremal_kraus} Let
$\Phi=\sum_{i,j}\sum_{k\in I(i,j)} V_kaV_k^*$ be a minimal Kraus representation.
 Then $\Phi$ is extremal if and only if the set
\[
\cup_{i,j}\{V_k^*V_l: k,l\in I(i,j)\} \cup\{x_{1},\dots,x_M\}
\]
is linearly independent.

\end{coro}

{\it Proof.} Let
$D\in (\Be\otimes \Ae)_P$ have the form
(\ref{eq:D}), then
\[
\ptr_\Be D= \sum_{i,j}\sum_{k,l\in I(i,j)} d_{k,l} (V_k^*V_l)^T
\]
By Theorem \ref{thm:extreme_gchan}, $X$ is extremal if and only if
\begin{equation}\label{eq:kraus}
\sum_{i,j}\sum_{k,l\in I(i,j)} d_{k,l} V_k^*V_l\in J^\perp
\end{equation}
 for some
coefficients $d_{k,l}$ implies that
all  $d_{k,l}=0$. This is exactly what we needed to prove.

\qed

Note that for ordinary channels, Corollary \ref{coro:extremal_kraus} gives the well known extremality condition obtained in \cite{choi}, namely that the set $\{V_kV_l^*, k,l\in I(i,j), i,j\}$  is linearly independent.

Let now $\Phi=\Lambda_q\circ\chi_c$ be  a minimal decomposition and let 
$\Lambda_q(a)=\sum_k W_kaW_k^*$ be a Kraus representation. 
As above, since $q\in \Ae$, we may suppose that for each $(i,j)$ there is some set of indices $J_W(i,j)$, such that for $k\in J_W(i,j)$, $W_k:\Ha_j\to \Ka_i$. 

\begin{coro} \label{coro:extremal_gchan_chan}
Let $\Phi=\Lambda_q\circ\chi_c$ be a minimal decomposition and let
$\Lambda_q(a)=\sum_{k}W_kaW_k^*$ be a  Kraus
representation.  Then $\Phi$ is
extremal  if and only if $\Lambda_q$ is an extremal channel and
\[
\mathrm{span}\{W_k^*W_l, k,l\in J_W(i,j), i,j \}\cap (cJc^*)^{\perp_q}=\{0\}
\]
where $L^{\perp_q}=L^\perp\cap \Ae_q$ for any subspace $L\subseteq \Ae_q$.
\end{coro}

{\it Proof.} By Theorem \ref{thm:extremal_decomp}, $\Phi$ is extremal if and only if $\Lambda_q$ is an extremal
generalized channel with respect to $cJc^*$.
Let $\Lambda_q(a)=\sum_{i,j}\sum_{k\in J_U(i,j)}U_kaU_k^*$ be a minimal
Kraus representation. Then Corollary \ref{coro:extremal_kraus} can be reformulated as follows:
$\Lambda_q$ is extremal in $\Ce_{cJc^*}(\Ae_q,\Be)$ if and only if the set
$\{U_k^*U_l, k,l\in J_U(i,j), i,j\}$
 is linearly independent (equivalently, $\Lambda_q$ is an extremal channel) and
\[
\mathrm{span}\{U_k^*U_l, k,l\in J_U(i,j), i,j\}\cap (cJc^*)^{\perp_q}=\{0\}.
\]
Given any  other Kraus
representation of $\Lambda_q$, then for every $(i,j)$ there are matrices 
$\{\mu^{i,j}_{r,k}\}_{r\in J_W(i,j),k\in J_U(i,j)}$ such that $\sum_{r\in J_W(i,j)}\bar\mu^{i,j}_{r,k}\mu^{i,j}_{r,l}=\delta_{k,l}$ for all
$k,l\in J_U(i,j)$ and $W_r=\sum_{k\in J_U(i,j)} \mu^{i,j}_{r,k}U_k$. It follows that
\[
\mathrm{span}\{U_k^*U_l, k,l\in J_U(i,j), i,j\}=\mathrm{span}\{W_k^*W_l, k,l \in J_W(i,j), i,j\}.
\]

\qed

Let us now return to the setting of Lemma \ref{lemma:hom}, so that $J=S^{-1}(\mathbb C \rho_0)$ with $\rho_0=S(\rho)$ for some invertible $\rho\in \Se(\Ae)$ and
$S^*:\Ae_0\to\Ae$ an injective homomorphism. Then $\Phi$ has a minimal decomposition of the form $\Phi=\Lambda_q\circ\chi_{d^{1/2}}$ and  by Lemma 
\ref{lemma:hom}, there is some $\sigma_0\in \Se(\Ae)$, such that 
 $d^{1/2}=S^*(b_0)$ with  $b_0$ the unique positive solution of 
 $b_0\rho_0b_0=\sigma_0$. Moreover, by  the proof of the Lemma, 
\[
d^{1/2}Jd^{1/2}=S^{-1}(\mathbb C\sigma_0)\subseteq \Ae_q,
\]
 where $q=\supp (d)=
S^*(\supp (b_0))=S^*(q_0)$ with $q_0:=\supp(\sigma_0)$.

\begin{coro}\label{coro:hom} In the above setting, $\Phi$ is extremal 
in $\Ce_J(\Ae,\Be)$ if and only if $\Lambda_q$ is extremal in $\Ce_{qJq}(\Ae_q,\Be)$ where  $qJq= S^{-1}(\mathbb Cq_0\rho_0q_0)$.
 In particular, if $q=I$ then $\Phi$ is extremal in $\Ce_J(\Ae,\Be)$ 
if and only if $\Lambda$  is.

\end{coro}

{\it Proof.} By Corollary \ref{coro:extremal_gchan_chan}, $\Phi$ is extremal if 
and only if $\Lambda_q$ is an extremal channel and 
\begin{equation}\label{eq:hom_chan}
\mathrm{span}\{W_k^*W_l, k,l\in J_W(i,j), i,j \}\cap 
S^*(\{\sigma_0\}^{\perp_{q_0}})=\{0\}
\end{equation}
for a Kraus representation of $\Lambda_q$.
Let $\{y_1,\dots,y_m\}$ be a basis of the subspace $\{\sigma_0\}^{\perp_{q_0}}$, 
then $\{q_0,y_1,\dots,y_m\}$ is a basis of $(\Ae_0)_{q_0}$. Suppose that
 (\ref{eq:hom_chan}) holds and let $c_{k,l}$ be some coefficients
 such that
\[
\sum_{i,j}\sum_{k,l\in J_W(i,j)}c_{k,l} W_k^*W_l = S^*( x)
 \]
 for some $x\in \{q_0\rho_0q_0\}^{\perp_{q_0}}$. Let $x=t_0q_0+\sum_jt_jy_j$. Then since $\Lambda_q$ is a channel, we have
$\sum_k W_k^*W_k=q=S^*(q_0)$ and
\[
\sum_{k,l} c_{k,l}W_k^*W_l-t_0\sum_k W_k^*W_k =S^*(\sum_jb_jy_j)\in
S^*(\{\sigma_0\}^{\perp_{q_0}})
\]
Hence  $\sum_{k,l} c_{k,l}W_k^*W_l-t_0\sum_k W_k^*W_k=0$, so that  $x=t_0q_0$. But this is possible only if  $x=0$, this implies that 
\begin{equation}\label{eq:hom_chan_2}
\mathrm{span}\{W_k^*W_l, k,l\in J_W(i,j), i,j \}\cap 
S^*(\{q_0\rho_0q_0\}^{\perp_{q_0}})=\{0\}
\end{equation}
Similarly, we can prove that (\ref{eq:hom_chan_2}) implies (\ref{eq:hom_chan}).
The statement now follows by Corollary \ref{coro:extremal_gchan_chan}.

\qed

\subsection{Extremality conditions for conjugate maps}

 Let $\Phi:\Ae\to \Be$ be a completely positive map with a minimal Kraus representation
 $\Phi(a)=\sum_{i,j}\sum_{k\in I(i,j)}V_kaV_k^*$ and let $n_{i,j}=|I(i,j)|$,
 $N=\sum_{i,j}n_{i,j}$.
Then  the
 conjugate map of $\Phi$ is defined as the  map $\Phi^C:\Ae \to \mathcal D:=\bigoplus_{i,j}B(\mathbb C^{n_{i,j}})$ such that
\[
\Phi^C(a)=\bigoplus_{i,j}\sum_{k,l\in I(i,j)} \Tr (V_kaV_l^*)|k\>\<l|
\]
where $\{|k\>\}_k$ is a basis of $\mathbb C^N$
such that  $\{|k\>, k\in I(i,j)\}$ is a  basis of $\mathbb C^{n_{i,j}}$. It is clear that $\Phi^C$ is completely positive 
 and, since $\Tr \Phi(a)=\Tr\Phi^C(a)$ for all $a\in \Ae$, $\Phi \in \Ce_J(\Ae,\Be)$
if and only if $\Phi^C\in \Ce_J(\Ae,\mathcal D)$.

\begin{coro}\label{coro:extremal_conjugate}
$\Phi$ is extremal in $\Ce_J(\Ae,\Be)$ if and only if the conjugate map $\Phi^C$ 
satisfies
\[
\Phi^C(J)=\mathcal D
\]

\end{coro}

{\it Proof.} Let $\Phi(a)=\sum_{i,j}\sum_{k,l\in I(i,j)} V_kaV_k^*$ be a minimal Kraus representation. Let $D=\oplus_{i,j}\sum_{k,l\in I(i,j)} d_{k,l}|e_k\>\<e_l|\in \mathcal D$ and $a\in \Ae$. Then
\[
\Tr D\Phi^C(a)=\Tr\left(\sum_{i,j}\sum_{k,l\in I(i,j)} d_{k,l}V_k^*V_l a\right)
\]
It follows that $\Tr D\Phi^C(a)=0$ for all $a\in J$ if and only if
$\sum_{i,j} \sum_{k,l\in I(i,j)} d_{k,l}V_k^*V_l\in J^\perp$.
From this and  Corollary \ref{coro:extremal_kraus}, we get that $\Phi$
is extremal if and only if $\Phi^C(J)^\perp=\{0\}$. Since $\Phi^C(J)$ is a subspace in $\mathcal D$, this is equivalent with $\Phi^C(J)=\mathcal D$.

\qed

Note that the condition $\Phi(J)=\Be$ means that the channel $J\to \Be$ defined by $\Phi$ is surjective.

\subsection{Extremal quantum supermaps}

Since $\Ce(\Be_0,\dots,\Be_n)=\frac 1{c_{n-1}} \Ce_{J_{n-1}}(\Ae_{n-1},\Be_n)$,
 extremality conditions for quantum supermaps can be obtained from the previous
sections. For this, we need  to describe the subspaces
$J_{n-1}^\perp$ and their bases. Let $\Te(\Be)$ denote the subspace of traceless elements in $\Be$.

\begin{lemma} \label{lemma:perp_basis} Let $J\subseteq \Ae$ be a subspace.
 Then
\[
(\ptr_\Be^{\Be\otimes \Ae})^{-1}(J)=(I_\Be\otimes J)\oplus (\Te(\Be)\otimes \Ae)
\]

\end{lemma}

{\it Proof.}  We have $\Te(\Be)=\{I_\Be\}^\perp$ and $\Be=\mathbb CI_\Be \oplus
\Te(\Be)$, $\Ae=J\oplus J^\perp$.
Hence
\[
\Be\otimes \Ae=(I_\Be\otimes J^\perp)\oplus(I_\Be\otimes J)\oplus (\Te(\Be)\otimes \Ae)
\]
Since $L^\perp=I_\Be\otimes J^\perp$ by (\ref{eq:perp}), we must have
$L=(I_\Be\otimes J)\oplus (\Te(\Be)\otimes \Ae)$.

\qed

Let $I_n=I_{\Be_n}$, $\Te_n:=\Te(\Be_n)$ and let
$\{t^n_1,\dots,t^n_{d_n-1}\}$ be some fixed basis of $\Te_n$,
$d_n:=\dim(\Be_n)$. Moreover, let $c_n:=\dim(\Ae_n)$ and let us fix a basis
$\{a^n_1,\dots,a^n_{c_n}\}$ of $\Ae_n$.

\begin{prop}\label{prop:subsp_jperp} Put  $\Ae_0=\Be_0$ and $\Ae_{-1}=\mathbb C$.
 Then
\[
J_n^\perp=I_n\otimes \bigvee_{j=0}^{\lfloor \frac{n-1}2\rfloor}\left[(\otimes_{l=1}^{2j} I_{n-l})
\otimes \Te_{n-2j-1}\otimes \Ae_{n-2j-2}\right]
\]
\end{prop}

{\it Proof.} We proceed by induction. For $n=1$, we have already seen in Section
\ref{sec:chanonchan} that
 $J_1^\perp=I_{\Be_1}\otimes \Te_0$. Suppose now that the statement holds for all $m<n$.
Note that by (\ref{eq:jodd}) and (\ref{eq:jeven}), $J_n=S_n^{-1}(I_{\Be_{n-1}}\otimes J_{n-2})$,
hence by (\ref{eq:perp}) and Lemma \ref{lemma:perp_basis},
\[
J_n^\perp=I_{\Be_n}\otimes S_{n-1}^{-1}(J_{n-2}^\perp)=I_{\Be_n}\otimes\left[
( I_{\Be_{n-1}}\otimes J_{n-2}^\perp)\oplus
(\Te_{n-1}\otimes \Ae_{n-2})\right]
\]
It is now easy to finish the proof.

\qed
It is clear that the set
\[
\bigcup_{j=0}^{\lfloor \frac {n-1}2\rfloor}\bigcup_{k=1}^{d_n-1}\bigcup_{l=1}^{c_n}
\left\{ I_{\Be_n}\otimes (\otimes_{l=1}^{2j} I_{\Be_{n-l}})\otimes t^{n-2j-1}_k\otimes
a^{n-2j-2}_l\right\}
\]
is a basis of $J_n^\perp$.

\begin{rem} Let  $J\subseteq \Ae$ and let $\Lin(L)$ be as in (\ref{eq:Lin}).
We have
by Lemma \ref{lemma:perp_basis} that
\[
\Lin(L)=(\Tr_\Be^{\Be\otimes\Ae})^{-1}((J^T)^\perp)=(I_\Be\otimes (J^T)^\perp)\oplus (\Te(\Be)\otimes \Ae)
\]
so that we can find a basis of $\Lin(L)$ as
\[
\mathbb B(\Lin(L)):=\bigcup_{i=1}^M\{I_\Be\otimes x_i^T\}\cup \bigcup_{j=1}^{d_\Be-1}\bigcup_{k=1}^{d_\Ae}
 \{t^\Be_j\otimes a_k'\}
\]
where $\{a_k'\}_k$ is any basis of $\Ae$, $\{x_1,\dots,x_M\}$ is a basis of $J^\perp$ and $\{t^\Be_j\}_j$ is a basis of $\Te(\Be)$. Theorem \ref{thm:extreme_gchan} can be
reformulated as follows: $X$ is extremal if and only if the set
\[
\{|v_k\>\<v_l|, k,l\in I(i,j), i,j\}\cup \mathbb B(\Lin(L))
\]
is linearly independent. Of course, we may chose any  other basis of $(\Be\otimes\Ae)_P$, moreover we may suppose that all the bases consist of self-adjoint
elements, since all involved subspaces are self-adjoint.
In the case of generalized quantum instruments, we get
exactly the extremality condition obtained in \cite{daria_extr}.
\end{rem}

\begin{rem} Note that an element 
$X\in \Ce(\Be_0,\dots,\Be_n)$ defines many completely positive maps
between different input and output spaces. For example a quantum $N$ comb
defines a map $B(\Ha_1\otimes \dots\otimes \Ha_{2N-2})\to B(\Ha_0\otimes \Ha_{2N_1})$, but the same element viewed as a quantum supermap defines a map 
$B(\Ha_0\otimes\dots\otimes \Ha_{2N-2})\to B(\Ha_{2N-1})$.
 The Kraus operators, as well as the conjugate
map, depend on the choice of this map. The conditions obtained above apply
 only for the case when the input space is $\Ae_{n-1}$.

\end{rem}

Let now $X\in \Ce(\Be_0,\Be_1,\Be_2)$ and
let $\Phi_X=\Lambda_q\circ \chi_{I_1\otimes \omega^{1/2}}$ be a minimal decomposition of
 $\Phi_X$ for some $\omega\in \Se(\Be_0)$.   Let $q_0:=\supp(\omega)$, so that 
$q=I_1\otimes q_0$. Then  Corollary \ref{coro:hom} 
implies:

\begin{coro}\label{coro:extremal_chan_chan}
$X$ is extremal in $\Ce(\Be_0,\Be_1,\Be_2)$ if and only if $(\Tr q_0)^{-1}
X_{\Lambda_q}$ is extremal
in $\Ce((\Be_0)_{q_0},\Be_1,\Be_2)$.
\end{coro}

In the case of 1-testers, this  result was obtained in \cite[Theorem 3]{daria_extr}.

\section{Extremal generalized POVMs}
\label{sec:POVM}
Let $K\subseteq \Se(\Ae)$ be any convex subset and let $U$ be a finite set,
 $|U|=m$.
A measurement on $K$ with values in $U$ is defined as an affine map
 from $K$ into the set $P(U)$ of probability measures on $U$. It was proved in
\cite{ja} that all such  measurements can be extended to  positive maps
 $\Ae\to \mathbb C^m$ if and only if $K$ is a section of $\Se(\Ae)$, that is,
 $K=J\cap \Se(\Ae)$ for $J=\mathrm{span}(K)$. 

Any positive map $\Phi: \Ae\to \mathbb C^m$ is given by an $m$-tuple $M=(M_1,\dots,M_m)$ of positive operators, such that $\Phi(a)_u=\Tr M_ua$ for all $a\in \Ae$
 and if $\Phi$ restricts to a measurement on $K$, we must have 
$\sum_u\Tr M_ua=1$ for all $a\in K$, so that 
\[
\sum_u M_u\in (I_\Ae+J^\perp)\cap \Ae^+.
\]
Any $m$-tuple of positive operators with this property is called a generalized 
POVM (with respect to $J$) and the set of all such generalized POVMs will be denoted by $\Me_J(\Ae,U)$.

It is quite clear that there is a one-to-one correspondence between $\Me_J(\Ae,U)$ and $\Ce_J(\Ae,\mathbb C^m)$, given by
\[
X=\sum_{u\in U} |u\>\<u|\otimes M_u^T,\qquad X\in \Ce_J(\Ae,\mathbb C^m),\ 
M\in \Me_J(\Ae,U)
\]
Hence  we can use 
the results of the previous section  to characterize the extreme points of 
$\Me_J(\Ae,U)$.

\begin{thm}\label{thm:extremal_gPOVM}
Let $M=\{M_u,u\in U\}\in \Me_J(\Ae,U)$
and let $p_u=\supp(M_u)$. Then $M$ is extremal in $\Me_J(\Ae,U)$ 
if and only if
for any collection $\{D_u, u\in U\}$ with $D_u\in \Ae_{p_u}$,
$\sum_u D_u\in J^\perp$ implies $D_u=0$ for all $u\in U$.

\end{thm}
{\it Proof.} By Theorem \ref{thm:extreme_gchan}.

\qed

\begin{coro} Let $p_u=\supp(M_u)$, $p=\supp(\sum_i M_i)$. If $M$ is extremal in 
$\Me_J(\Ae,U)$, 
then $\sum_u\dim(\Ae_{p_u})\le
\dim(pJp)$.

\end{coro}

{\it Proof.} By Corollary \ref{coro:gchan_dim}.

\qed

As a generalized channel,
a generalized POVM $M=\{M_u, u\in U\}$ has a minimal decomposition
$M=\Lambda_q\circ\chi_c$, where
$c^*c=\sum_u M_u$, $q=\supp(cc^*)$ and 
$\Lambda_q=\{\Lambda_u, u\in U\}$ is a POVM in $\Ae_q$.

\begin{thm}\label{thm:extremal_decomp_gPOVM}
$M$ is  extremal in $\Me_J(\Ae,U)$ and only if $\Lambda_q$ is extremal in
\newline
$\Me_{cJc^*}(\Ae_q,U)$.

\end{thm}

{\it Proof.} By Theorem \ref{thm:extremal_decomp}.

\qed

\subsection{Extremality of PVM's in $\Me_J(\Ae,U)$}

It is well known that any projection-valued measure (PVM) is extremal in the set
 of POVMs. 
However, the set of generalized POVMs is larger than the set of POVMs, and an 
additional condition is needed for extremality of a PVM in $\Me_J(\Ae,U)$.

Let $M$ be a PVM. In this case, the range of $M$ 
generates an abelian  subalgebra in $\Ae$. We denote
 by $\{M\}'$ the commutant of this subalgebra, that is, the set of all $b\in B(\Ha)$, such that $bM_u=M_ub$ for all
 $u\in U$.

\begin{prop}\label{prop:gpovm_PVM}
Let $M$ be a PVM. 
Then $M$ is extremal in $\Me_J(\Ae,U)$ if and only if
\[
\{M\}'\cap J^\perp=\{0\}
\]

\end{prop}

{\it Proof.}
If $M$ is a PVM, then $\supp(M_u)=M_u$ and for $D\in \Ae$,
$D\in \{M\}'$ if and only if $D=\sum_uD_u$ with $D_u=DM_u\in \Ae_{M_u}$.
In this case, $D=0$ if and only if $D_u=0$ for all $u$.
The statement now follows from  Theorem \ref{thm:extremal_gPOVM}.

\qed

Note that combining Proposition \ref{prop:gpovm_PVM} with Theorem 
\ref{thm:extremal_decomp_gPOVM} gives a  characterization 
of extremal  generalized POVMs with two outcomes. Indeed, if $M$ is an extremal
2-outcome generalized POVM with minimal decomposition 
$M=\Lambda_q\circ \chi_c$, then 
$\Lambda_q$ must be an extremal 2-outcome  POVM, hence a PVM.

\begin{coro} Let $U=\{1,2\}$ and let $M\in \Me_J(\Ae,U)$ with a minimal 
decomposition $M=\Lambda_q\circ\chi_c$. Then $M$ is extremal if and only if 
$\Lambda_q$ is a PVM and 
\[
\{\Lambda_q\}'\cap (cJc^*)^{\perp_q}=\{0\}
\]
\end{coro}

Next we look at the condition in Proposition \ref{prop:gpovm_PVM} in the setting 
of Lemma \ref{lemma:hom}.  

\begin{lemma}\label{lemma:hom_extr} Let $J=S^{-1}(\mathbb C\rho_0)$, where $S^*$ is an injective homomorphism and $\rho_0=S(\rho)$ for some invertible $\rho\in \Se(\Ae)$. Let $M$ be a 
PVM, then $M$ is extremal in $\Me_J(\Ae,U)$ if and only if 
$S^*(p_0)\in \{M\}'$ for some projection $p_0\in \Ae_0$ implies that $p_0=0$ or $I$.
\end{lemma}

{\it Proof.} Let $0\ne x \in \{M\}'\cap J^\perp$, then $x=S^*(y)$ for some 
$y\in \Ae_0$ with $\Tr y\rho_0=0$ and we may suppose $y=y^*$. Let $y=y_+-y_-$ be 
the decomposition of $y$
 into its positive and negative part, then $x_\pm=S^*(y_\pm)$ is the 
decomposition of $x$. Let $p_+=\supp(y_+)$, then $S^*(p_+)=\supp x_+\in \{M\}'$.
Moreover, since $\rho_0$ is invertible, we must have $y_\pm\ne 0$, hence $p_+\ne 0,I$.

Conversely, suppose $S^*(p_0)\in \{M\}'$ for some $p_0\ne 0,I$ and let $t=\Tr\rho_0p_0$. Put $y:=p_0-\tfrac{t}{1-t}(I-p_0)$, then $y\ne 0$ and $S^*(y)\in \{M\}'\cap J^\perp$.

\qed

\begin{ex} Let $K=Diag_\lambda$ as in Example \ref{ex:diag} and let 
$M\in \Me_{J_\lambda}(M_n(\mathbb C),U)$. Then  $M$ has a minimal decomposition of the form $M=\Lambda_q\circ \chi_{d^{1/2}}$ with $d=\sum_id_i|i\>\<i|$, $d_i=\mu_i/\lambda_i$ for some probability vector $\mu$.
Let $I=\{i, d_i>0\}$, then $q=p_I:=\sum_{i\in I} |i\>\<i|$. Let us denote 
by $M_I(\mathbb C)$ the subalgebra in $M_n(\mathbb C)$ generated by the matrix 
units $\{|i\>\<j|,\ i,j\in I\}$, then $\Lambda_q$ is a POVM on $M_I(\mathbb C)$.
By Corollary \ref{coro:hom},
 $M$ is extremal if and only if $\Lambda_q$ is extremal in 
$\Me_{p_IJp_I}(M_{I}(\mathbb C),U)$.

Suppose that $\Lambda_q$ is a PVM. Then $M$ is extremal if and only if there is 
 no nonempty $J\subsetneq I$ such that $p_J\in \{\Lambda_q\}'$.

\end{ex}

\subsection{Extremal 1-testers}

An $n$-tester on $(\Ha_0,\dots,\Ha_{2n-1})$  is a map that defines a measurement on the set of $n$-combs on some
finite dimensional Hilbert spaces $\Ha_0,\dots,\Ha_{2n-1}$, hence it is a (constant multiple of) a  generalized
POVM with respect to $J_{2n-1}(\Ha_0,\dots,\Ha_{2n-1})$. As generalized channels, $n$-testers are elements
 of $\Ce(\Ha_0,\dots,\Ha_{2n-1},\mathbb C^m)$.

 In particular, 1-testers define measurements on the set of channels $\Ce(\Ha_0,\Ha_1)$. The 1-testers  were introduced in  \cite{daria_testers}, note that these
  appeared independently in \cite{ziman}, under the name PPOVMs (process POVMs). Hence a 1-tester is a collection
  of operators $M=\{M_u,u\in U\}$, $M_u\in B(\Ha_1\otimes \Ha_0)^+$, with $\sum_u M_u= I_{\Ha_1}\otimes \omega$
   for some $\omega\in \Se(\Ha_0)$.
It follows that $M$ has a minimal decomposition of the form
$M=\Lambda\circ\chi_{I_1\otimes \omega^{1/2}}$ and $\Lambda$ is a 
POVM on 
$B(\Ha_1\otimes q_0\Ha_0)$, with $q_0=\supp(\omega)$. Moreover, by 
 Corollary \ref{coro:extremal_chan_chan}, $M$ is extremal 
if and only if $\Tr q_0^{-1}\Lambda$ is an extremal 1-tester on 
$(q_0\Ha_0,\Ha_1)$.

Let us summarize the results of the previous section for 1-testers.

\begin{prop}\label{prop:extremal_1_tester}
\begin{enumerate} \item[(i)]  Let $p_u=\supp (M_u)$. Then $M$ is an extremal 1-tester on $(\Ha_0,\Ha_1)$ if and only if 
$D_u\in B(p_u(\Ha_1\otimes\Ha_0))$, $\sum_u D_u =I_{\Ha_1}\otimes x$ with $\Tr x=0$ implies 
$D_u=0$ for all $u$.

\item[(ii)] Suppose that $\Lambda$ is a PVM.
Then $M$ is  extremal if and only if $I_{\Ha_1}\otimes p\in \{\Lambda\}'$ for
 some projection $p$ on $q_0\Ha_0$ implies $p=0$ or $p=q_0$. 
\item[(iii)] Let  
$U=\{1,2\}$, then $M$ ix extremal if and only if $\Lambda=(\Lambda_1,\Lambda_2)$ is a PVM and 
$\Lambda_1$ commutes with no projection of the form 
$I_{\Ha_1}\otimes p$ with $p\neq 0,q_0$. 
\end{enumerate}
\end{prop}

\begin{ex}\label{ex:extremal_qubit}(\textbf{ Extremal 2-outcome qubit 1-testers})
Extremal 1-testers with 2 outcomes for  $\dim(\Ha_1)=\dim(\Ha_0)=2$
 were described in \cite{daria_extr}. We show that using our results, this  is very easily achieved
 and extended to the case $\dim(\Ha_1)=n\ge 2$.

So let $M=(M_1,M_2)$ be a 1-tester on $(\Ha_0,\Ha_1)$ with $\dim(\Ha_0)=2$ and let $M_1+M_2=I_1\otimes \omega$.
Suppose first that $\rank(\omega)=1$, then $\omega=|\varphi\>\<\varphi|$ for some $\varphi\in \Ha_0$ and then
$M_u=\Lambda_u=N_u\otimes|\varphi\>\<\varphi|$ for some POVM $N$ on $B(\Ha_1)$.
Since any subspace in a one-dimensional space is trivial, $M$ is extremal if and
only if $\Lambda$ (and hence also $M$) is a  PVM.

Suppose $\rank(\omega)=2$, then $q_0=I_0$. Any nontrivial projection $p$ 
on $\Ha_0$ is rank one, $p=|\psi_0\>\<\psi_0|$. 
Hence 
$M$ is extremal if and only if  $\Lambda$ is a PVM, such that $\Lambda_1$ is not 
of the form
\[
\Lambda_1 =e\otimes |\psi_0\>\<\psi_0| + f\otimes |\psi^\perp_0\>\<\psi_0^\perp|
\]
where  $\psi_0,\psi_0^\perp\in \Ha_0$ are unit vectors such that 
 $\<\psi_0,\psi_0^\perp\>=0$ and $e,f$ are projections in $B(\Ha_1)$.

\qed
\end{ex}

\section{Concluding remarks}

We have obtained a set of extremality conditions for generalized channels. These
 conditions are not easily checked, in fact, in the case of quantum supermaps the conditions are quite complicated. However, for the case of a generalized POVM 
 consisting of a PVM combined with a simple generalized channel, we found a simpler condition. 

As already mentioned in the Introduction, there is no one-to-one correspondence 
between generalized channels and channels defined on the subspace $J$. 
In particular,  measurements on a section of the state space correspond to 
equivalence classes of generalized POVMs, because in general, there are many 
generalized POVMs giving the same probabilities for all elements in the section.
Therefore, extremal generalized POVMs do not necessarily correspond to extremal 
measurements on sections. The question of extremal measurements will be addressed in a forthcoming paper.

\end{document}